\documentclass[
    ,final            % use final for the camera ready runs
  ]
  {aipproc}

\layoutstyle{6x9}

\newcommand{\Pu}{p_{\uparrow}}

\newcommand{\Nu}{n_{\uparrow}}
\newcommand{\Nd}{n_{\downarrow}}

\begin{document}

\title{Exploring Three Nucleon Forces in Lattice QCD}



\author{Takumi Doi (for HAL QCD Collaboration)}{
  address={
Graduate School of Pure and Applied Sciences, University of Tsukuba,
Tsukuba 305-8571, Japan}
}

%\author{<author2>}{
%  address={<common address for author2 and author3>}
%}

%\author{<author3>}{
%  address={<common address for author2 and author3>}
%  ,altaddress={<author1 address>} % additional visiting address
%}

\begin{abstract}
We study the three nucleon force in $N_f=2$ dynamical clover fermion lattice QCD,
utilizing the Nambu-Bethe-Salpeter wave function of the three nucleon system.
Since parity-odd two nucleon potentials are not available in lattice QCD at this moment,
we develop a new formulation to extract the genuine three nucleon force
which requires only the information of parity-even two nucleon potentials.
In order to handle the extremely expensive calculation cost,
we consider a specific three-dimensional coordinate configuration for the three nucleons.
We find that the linear setup is advantageous, where 
nucleons are aligned linearly with equal spacings.
The lattice calculation is performed with 
$16^3\times 32$ configurations at $\beta=1.95$, $m_\pi = 1.13$GeV 
generated by CP-PACS Collaboration,
and the result of the three nucleon force in triton channel is presented.

\end{abstract}

\maketitle

%%%%%%%%%%%%%%%%%%%%%%%%%%%%%%%%%%%%%%%%%%%%
%% MAINMATTER
%%%%%%%%%%%%%%%%%%%%%%%%%%%%%%%%%%%%%%%%%%%%

\section{Introduction}
\label{sec:intro}

Understanding the interaction among nucleons is one of the 
most important issues in particle and nuclear physics.
In particular, the properties of the three nucleon force (TNF)
are the forefront of not only particle/nuclear physics but also 
astro physics.
For the binding energies of light nuclei, 
the two nucleon (2N) force alone underestimates the experimental data,
and attractive TNF components are required~\cite{Pieper:2007ax}.
On the other hand,
repulsive TNF components are also necessary 
in both of symmetric and asymmetric nuclear matters,
in order to reproduce the empirical saturation density of nuclei
in the former case,
and to make the equation of state % (EoS) 
of asymmetric nuclear matter
hard enough to sustain the observed massive neutron star masses
in the latter case~\cite{Akmal:1998cf}.
Recently, 
it is argued that TNF is responsible for 
the anomaly in the drip line and the nontrivial 
magic number of neutron-rich nuclei~\cite{Otsuka:2009cs, Holt:2010yb},
and could impact on the supernova nucleosynthesis.

Pioneered by Fujita-Miyazawa~\cite{Fujita:1957zz},
TNF have been commonly studied from the two-pion exchange models
with the $\Delta$-excitation.
However, since TNF is originated by the fact that
a nucleon is not a fundamental particle,
it is desirable to study TNF
from the fundamental degrees of freedom (DoF), i.e., quarks and gluons.
In this report, 
we carry out first-principle calculations of TNF
using the lattice QCD technique.

In the lattice QCD calculation of nuclear forces,
it is recently proposed~\cite{Ishii:2006ec} to use
the Nambu-Bethe-Salpeter (NBS) wave function 
so that the % calculated 
potential is faithful to the phase shift
by construction.
The obtained (parity-even) 2N potentials % from this method
are found to have desirable features,
such as attractive well at long and medium
distances, and the central repulsive core at short distance%
~\cite{Ishii:2006ec, Aoki:2009ji}.
It is also examined that the non-locality of the obtained potential
is small, and the leading terms in the derivative expansion
dominate the potential at low energies~\cite{Murano:2011nz}.
The method has been successfully extended to the
hyperon-nucleon (YN) and hyperon-hyperon (YY) interactions%
~\cite{Nemura:2008sp, Inoue:2010hs, Sasaki:2010bi}
including H-dibaryon system~\cite{Inoue:2010es}.
Further applications are given in 
Refs.~\cite{Ikeda:2010sg, Ikeda:2011bs, Takahashi:2009ef, Kawanai:2010ev}.
In this report, we discuss the extension of this approach 
to the three nucleon (3N) system, 
and lattice QCD calculation of TNF 
in the triton channel is carried out.
A preliminary account of this study is also given in Ref.~\cite{Doi:2010yh}.

\section{Formalism}
\label{sec:formulation}

Since the detailed formulation for the 2N force is 
given in Ref.~\cite{Aoki:2009ji},
we discuss the extension to the 3N system.
We consider the equal-time NBS wave 
function $\psi(\vec{r'},\vec{\rho})$,
which can be obtained through the six point correlator,
\begin{eqnarray}
G_{\alpha\beta\gamma,\alpha'\beta'\gamma'} (\vec{r'},\vec{\rho},t-t_0)
= \langle
           N_\alpha(\vec{x}_1,t) N_\beta (\vec{x}_2,t) N_\gamma (\vec{x}_3,t) \
\overline{(N_{\alpha'}(t_0)       N_{\beta'}(t_0)        N_{\gamma'}(t_0))}
\rangle ,
\end{eqnarray}
where
$\vec{r'} \equiv \vec{x}_1 - \vec{x}_2$, %and
$\vec{\rho} \equiv \vec{x}_3 - (\vec{x}_1 + \vec{x}_2)/2$
are the Jacobi coordinates,
and $N$ denotes 
either of %proton or neutron.
$p$ or $n$.
With the derivative expansion of the potentials,
the NBS wave function can be converted to the potentials
through the following % effective 
Schr\"odinger equation,
\begin{eqnarray}
%
%\lefteqn{
\biggl[ 
%- \frac{1}{2m_N} (\nabla^2_{x_1} + \nabla^2_{x_2} + \nabla^2_{x_3})
- \frac{1}{2\mu_{r'}} \nabla^2_{r'} - \frac{1}{2\mu_\rho} \nabla^2_{\rho} 
%} \\
%&&
+ \sum_{i<j} V_{2N,ij} (\vec{x}_{ij})
+ V_{TNF} (\vec{r'}, \vec{\rho})
\biggr] \psi(\vec{r'}, \vec{\rho})
= E \psi(\vec{r'}, \vec{\rho}) , 
\label{eq:Sch_3N}
\end{eqnarray}
where
$V_{2N,ij}(\vec{x}_{ij})$ with $\vec{x}_{ij} \equiv \vec{x}_i - \vec{x}_j$
denotes the potential between $(i,j)$-pair,
$V_{TNF}(\vec{r'},\vec{\rho})$ the TNF,
$\mu_{r'} = m_N/2$, $\mu_\rho = 2m_N/3$ the reduced masses.
If we calculate 
$\psi(\vec{r'}, \vec{\rho})$ for all $\vec{r'}$ and $\vec{\rho}$,
and if all $V_{2N,ij}(\vec{x}_{ij})$ are available 
by (separate) lattice calculations for the genuine 2N system,
we can extract $V_{TNF}(\vec{r'},\vec{\rho})$ through Eq.~(\ref{eq:Sch_3N}).

In practice, however, the calculation cost of $\psi(\vec{r'}, \vec{\rho})$
is prohibitively expensive.
Note that the number of diagrams generated by Wick contractions
tends to diverge with a factor of $N_u ! \times N_d !$
($N_{u,d}$ are numbers of u,d quarks in the system).
We develop several techniques to reduce the calculation cost, 
e.g., we take advantage of symmetries (such as isospin symmetry),
and
employ the non-relativistic limit for the source nucleon operator.
Similar techniques are
(independently) developed 
in the calculations of binding 
energies of triton/helium nuclei~\cite{Yamazaki:2009ua}.
Yet, the calculation cost remains quite expensive,
so we consider Eq.~(\ref{eq:Sch_3N}) 
with fixed three-dimensional (3D) coordinates of $\vec{r'}$ and $\vec{\rho}$.

To proceed, we find that 
the linear setup with $\vec{\rho}=\vec{0}$ is advantageous,
where three nucleons are aligned linearly with equal spacings of 
$r \equiv |\vec{r}| \equiv 1/2 |\vec{r'}|$.
In fact, 
because of $\vec{\rho}=\vec{0}$, the third nucleon is attached
to $(1,2)$-nucleon pair with only S-wave.
Considering the total 3N quantum numbers of $I=1/2, J^P=1/2^+$ (triton channel), 
the wave function can be completely spanned by
only three bases, which can be labeled
by the quantum numbers of $(1,2)$-pair as
$^1S_0$, $^3S_1$, $^3D_1$.
Therefore, the Schr\"odinger equation
leads to 
the $3\times 3$ coupled channel equations
with the bases of 
$\psi_{^1S_0}$, $\psi_{^3S_1}$, $\psi_{^3D_1}$.
The reduction of the dimension of bases 
%in this way
is expected to improve the S/N as well.

Even when we calculate with the fixed 3D-configuration (with the linear setup or whatever),
the identification of genuine TNF remains nontrivial.
This is because 
the parity-odd 2N potentials are not available
in lattice QCD at this moment,
and we cannot subtract the contribution of $V_{2N}$ unambiguously.
Note that although the total parity of the 3N system can be projected, % (to be positive),
a 2N-pair inside the 3N system 
could be either of positive or negative parity.
%This issue could be quite serious, since TNF is much smaller than 2N.
The familiar procedure of partial wave expansion cannot be performed here,
since we calculate only fixed 3D-configurations to reduce the calculation cost.

To overcome this problem,
%In the effort to overcome this issue,
we find that the following channel in 
the triton is useful,
\begin{eqnarray}
\psi_S &\equiv&
\frac{1}{\sqrt{6}}
\Big[
-   \Pu \Nu \Nd + \Pu \Nd \Nu               
                - \Nu \Nd \Pu + \Nd \Nu \Pu 
+   \Nu \Pu \Nd               - \Nd \Pu \Nu
\Big]  .
\label{eq:psi_S}
\end{eqnarray}
This wave function itself has been well known,
but the point here is that 
it is anti-symmetric
in spin/isospin spaces 
%\underline{for any 2N-pair}.
for any 2N-pair.
Combined with the Pauli-principle,
it is automatically guaranteed that
any 2N-pair couples with even parity only.
Therefore, we can extract TNF unambiguously in this channel,
without the information of parity-odd 2N potentials.
Note that no assumption on the choice of 3D-configuration 
is imposed in this argument,
and we can take advantage of this feature
in future TNF calculations with 
3D-configurations other than the linear setup.

\section{Lattice QCD results}
\label{sec:results}

We employ %the 
$N_f=2$ dynamical 
clover fermion 
configurations
generated by CP-PACS Collaboration~\cite{Ali Khan:2001tx}.
We use %$(\beta,c_{sw})=(1.83,1.761)$
598 configurations at
$\beta=1.95$
%$c_{sw}=1.53$ 
%configurations 
with the lattice size of $L^3 \times T = 16^3\times 32$,
which corresponds to
$(2.5{\rm fm})^3$ box in physical spacial size
with 
the lattice spacing of 
$a^{-1} = 1.269 {\rm GeV}$. %~\cite{conf:cp-pacs}.
We calculate at the hopping parameter of $u$, $d$ quarks
$\kappa_{ud} = 0.13750$, which corresponds to 
$m_\pi = 1.13$ GeV, $m_N = 2.15$ GeV.
We use the wall quark source with Coulomb gauge fixing.
In order to enhance the statistics, we perform 
the calculation for sources on 16 time slices
for each configuration.

We perform the calculation
with the linear setup at ten physical points of the distance $r$.
Since we have only one channel (Eq.~(\ref{eq:psi_S})) 
which is free from parity-odd 2N potentials,
we can determine one type of TNF.
In this report,
we consider the scalar/isoscalar type TNF.
%
%In fact, 
In the Urbana IX model~\cite{Pudliner:1995wk},
which is an often-used phenomenological TNF,
TNF consists of two parts,
one is the two-pion exchange TNF %(which has spin/isospin dependencies),
and the other is phenomenologically introduced 
scalar/isoscalar repulsive TNF.
Since the two-pion exchange TNF is 
expected to be suppressed in the current lattice setup 
of $m_\pi = 1.13$ GeV,
it is reasonable to assume the scalar/isoscalar nature 
for TNF here.

\begin{figure}[t]
%\vspace*{-15mm}
\begin{minipage}{0.45\textwidth}
\begin{center}
\hspace*{-8mm}
\includegraphics[width=0.8\textwidth,angle=270]{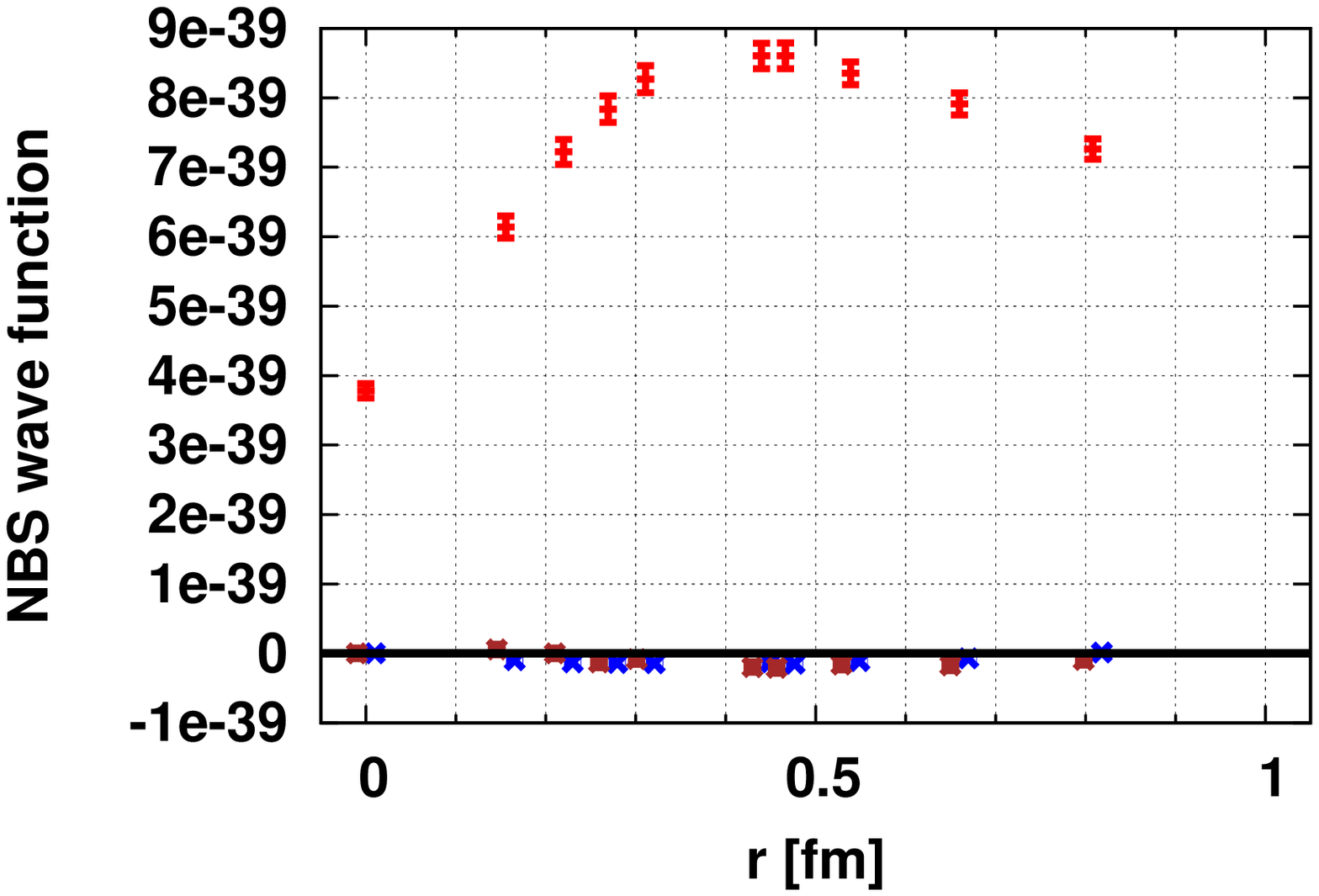}
%
%\caption{
%\label{fig:wf}
%The wave function with linear setup in the triton channel.
%Red, blue, brown points correspond to
%$\psi_S$, $\psi_M$, $\psi_{^3D_1}$, respectively.
%}
%
\end{center}
\end{minipage}
\hfill
\begin{minipage}{0.45\textwidth}
\begin{center}
%
%\hspace*{-12mm}
\includegraphics[width=0.8\textwidth,angle=270]{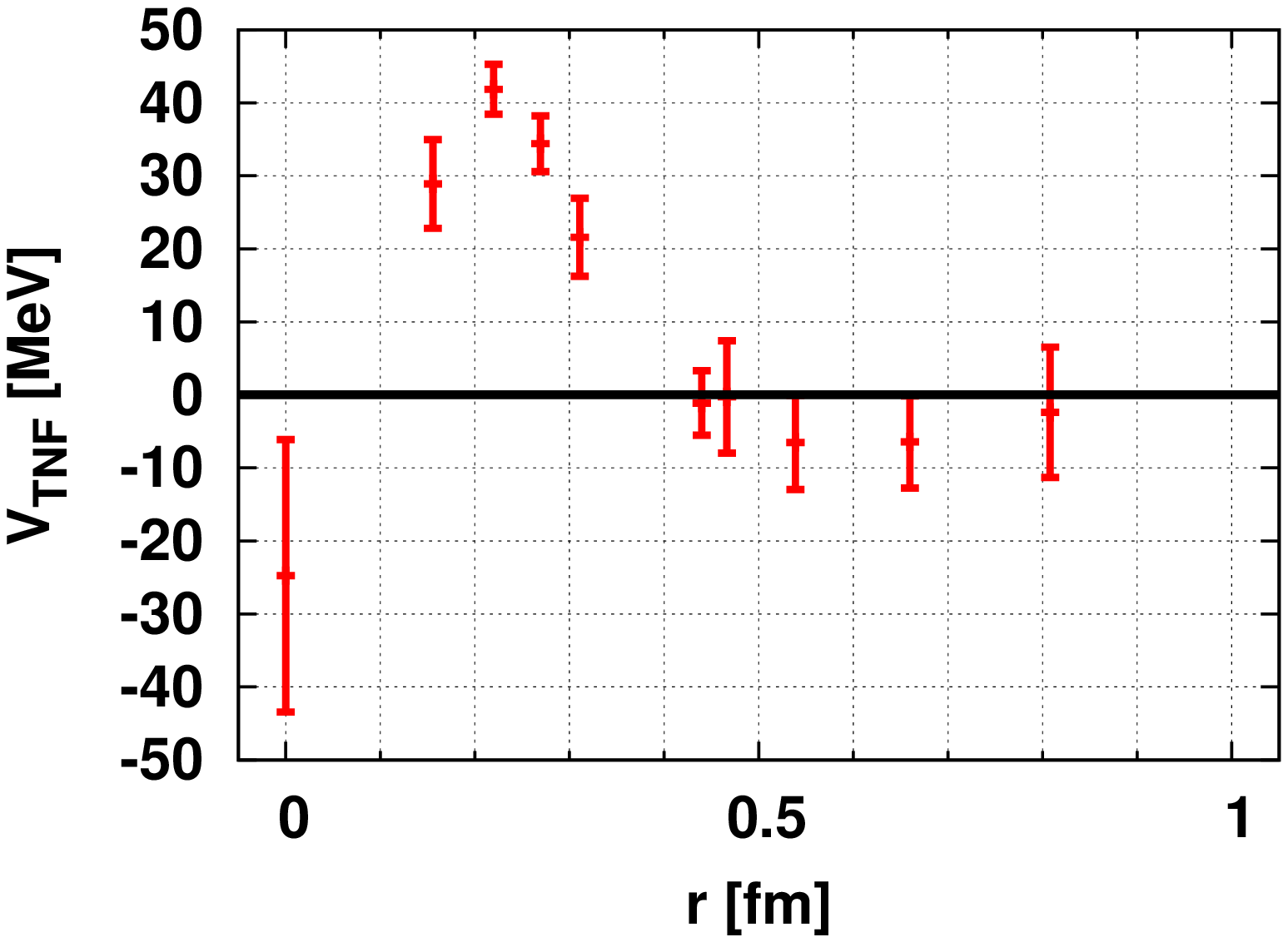}
\caption{
\label{fig:3N}
(color online).
(Left) The wave function with linear setup in the triton channel.
Red, blue, brown points correspond to
$\psi_S$, $\psi_M$, $\psi_{^3D_1}$, respectively.
(Right) The scalar/isoscalar TNF in the triton channel,
plotted against 
%in terms of 
the distance $r$ in the linear setup.
}
\end{center}
\end{minipage}
%\vspace*{-2mm}
\end{figure}

In Fig.~\ref{fig:3N} (left),
we plot each wave function of
$\psi_S = ( - \psi_{^1S_0} + \psi_{^3S_1} )/\sqrt{2}$,
$\psi_M \equiv ( \psi_{^1S_0} + \psi_{^3S_1} )/\sqrt{2}$
and
$\psi_{^3D_1}$ 
in the triton channel at $t-t_0 = 8$.
We observe that 
$\psi_S$ dominates the wave function.
This is because 
$\psi_S$ contains the component for which
all three nucleons are in S-wave.
%We observe good S/N, which is quite nontrivial.
%the practical statistical sampling number becomes much smaller
%by fixing the 3D-configuration.
%
%
By solving the $3\times 3$ coupled channels and 
subtracting the $V_{2N}$ % in Eq.~(\ref{eq:Sch_3N})
from the total potentials, % in the 3N system,
we determine TNF.
In Fig.~\ref{fig:3N} (right), we plot the results
for the scalar/isoscalar TNF. % against the distance $r$.
Here, the $r$-independent shift by energies is not included,
and thus about ${\cal O}(10)$ MeV systematic error is understood.
%There are various physical implications in Fig.~\ref{fig:TNR}.
At the long distance region of $r$, TNF is small as is 
expected.
At the short distance region, 
we observe the indication of repulsive TNF.
Recalling that the repulsive short-range TNF is phenomenologically required 
in high density matter,
this is very encouraging result.
Of course, we note that
further researches are necessary,
e.g., 
the examination of the discretization error
and 
the dependence on the quark masses.
%and the improvement of the signal statistically.
%
%
Studies along this line is currently underway.

%%%%%%%%%%%%%%%%%%%%%%%%%%%%%%%%%%%%%%%%%%%%%%%%
%% BACKMATTER
%%%%%%%%%%%%%%%%%%%%%%%%%%%%%%%%%%%%%%%%%%%%%%%%

%\begin{theacknowledgments}
%
We thank authors and maintainers of  CPS++\cite{CPS}, 
%a modified version of which is used in this study. 
%We also thank  CP-PACS Collaborations
and CP-PACS Collaboration
and ILDG/JLDG~\cite{conf:ildg/jldg} for providing gauge configurations.
The numerical calculations have been performed
on T2K at University of Tsukuba (09a-11, 10a-19)
and Blue Gene/L at KEK (09-23, 09/10-24).
This research is supported in part by 
MEXT Grant-in-Aid,
Scientific Research on Innovative Areas (20105001, 20105003, 21105515).
T. D. is supported in part by Grant-in-Aid for JSPS Fellows 21$\cdot$5985.

\end{document}